\documentclass[aps,pra,floatfix,twocolumn,noshowpacs]{revtex4}

\usepackage{graphicx}% Include figure files
\usepackage{graphics}% Include figure files
\usepackage{dcolumn}% Align table columns on decimal point
\usepackage{bm}% bold math
\usepackage[mathlines]{lineno}% Enable numbering of text and display math
\usepackage{amsmath} 
\usepackage{amssymb}

\usepackage{color}%

\begin{document}
%  \title{Supplemental Material for ``Out-of-plane potential for graphene and other sp$^2$ carbon structures'' }
\title{A Torsional potential for graphene derived from fitting to DFT results}
  %Supplemental Material«}

  \author{Georgios D. Chatzidakis}
  \affiliation{Department of Physics, National Technical University of Athens, GR-15780 Athens, Greece }
  \author{George~Kalosakas}
  \affiliation{Materials Science Department, University of Patras, Rio GR-26504, Greece}
  \affiliation{Crete Center for Quantum Complexity and Nanotechnology (CCQCN), Physics Department, University of Crete GR-71003 Heraklion, Greece}
  \author{Zacharias G. Fthenakis}
  \affiliation{Institute of Electronic Structure and Laser, FORTH, Heraklion, Greece}
  \affiliation{Department of Physics, University of South Florida, Tampa, Florida 33620, USA}
  \author{Nektarios N. Lathiotakis}
  \affiliation{Theoretical and Physical Chemistry Institute, National Hellenic
  Research Foundation, Vass. Constantinou 48, GR-11635 Athens, Greece  }

  \date{\today}
  
\begin{abstract}
We present a simple torsional potential for graphene to accurately describe its out-of-plane deformations.
The parameters of the potential are derived through appropriate fitting with suitable DFT calculations
regarding the deformation energy of graphene sheets folded around two different folding axes, along
an armchair or along a zig-zag direction. Removing the energetic contribution of bending angles,
using a previously introduced angle bending potential, we isolate the purely torsional deformation
energy, which is then fitted to simple torsional force fields. The presented out-of-plane torsional potential
can accurately fit the deformation energy for relatively large torsional angles up to 0.5 rad.
To test our proposed potential, we apply it to the problem of the vertical displacement of a single
carbon atom out of the graphene plane and compare the obtained deformation energy with
corresponding DFT calculations. The dependence of the deformation energy on the vertical displacement
of the pulled carbon atom is indistinguishable in these two cases, for displacements up to about 0.5 $\AA$.
The presented potential is applicable to other sp$^2$ carbon structures.
\end{abstract}
\maketitle

\section{Introduction}

Following the isolation of single layer graphene \cite{Novoselov666} an enormous research effort has been
devoted to the study of this two-dimensional material and its properties \cite{castroneto,synthesis,roadmap,elasticRev}.
Potential applications have been explored in electronics~\cite{avouris}, opto-electronics~\cite{Kusmartsev2015},
gas filtering~\cite{Celebi289}, energy storage~\cite{batteries}, uses related to its unique mechanical 
properties \cite{Lee385,SMLL,kalosakas}, etc.

Many empirical force fields have been used in atomistic simulations, calculating various
structural, mechanical or phonon properties of graphene
\cite{fasolinoNM,fasolino,bending,buechler,peeters1,peeters2,indent,campbell,arisRSC,koukaras,aris2d}.
Besides the older, well known Tersoff \cite{Tersoff,Tersoff1} and Brenner~\cite{Brenner} potentials, more
accurate force fields have been introduced the last two decades. For example, optimized parameter sets for the
latter potentials, providing better description of structural and phonon properties of graphene are presented
in Ref.~\cite{T2010}. LCBOP \cite{LCBOP,LCBOPII} and AIREBO~\cite{AIREBO} are efficient potentials
that have been widely applied in many calculations. Other potentials leading to good predictions of elastic
and thermal properties of graphene have been also discussed~\cite{Wei}.

More recently, we have presented simple analytical expressions for the accurate description of bond stretching and
angle bending potentials of graphene~\cite{kalosakas}. These potentials are derived by fitting analytical functions
to the deformation energy of proper distortions of graphene, obtained through accurate calculations
from first principles' methods (DFT). The presented force field is applicable only to distortions restricted
within the plane of graphene. These in-plane potentials can accurately describe elastic properties and the
mechanical response of graphene in various extensional loads~\cite{kalosakas}.
In this work, using similar ideas and methods, we extend this force field with torsional
energy terms, in order to be able to describe out-of-plane distortions in graphene. 
The basic motivation is to provide a simple and computational efficient classical potential which can be used
for accurate large-scale atomistic calculations. The torsional potential
presented here is also capable to describe other non-planar sp$^2$ carbon systems, like fullerenes
and carbon nanotubes~\cite{allo}.

In the present work, we describe in detail the procedure followed and the necessary analytical calculations in order to fit the proposed
torsional potential to ab-initio data. 
The full potential is then tested in the case of the deformation energy due to the vertical displacement of a C atom outside graphene's plane. A more comprehensive benchmark study for fullerenes, nanotubes and graphene's phonons is presented elsewhere~\cite{allo}.
We have considered two types of folding of graphene sheets around
different axes (either an armchair or a zig-zag one). The corresponding deformation energies are
calculated using DFT methods. Following the removal of the contribution of angle bending terms in the
total deformation energy, we isolated the pure torsional energy. Then, the analytic modeling of this energy in terms of 
individual torsional contributions, leads to a fitting procedure providing the optimal parameters of the
out-of-plane torsional energy.

This paper is organized as follows: In Sect.~\ref{sec:struc} we describe the structures and methodology adopted for the
DFT simulations that was used to obtain the deformation energies. Then in Sect.~\ref{sec:bend_rem}, we
present the analytic work for removing the angle bending contributions from the deformation energies. 
The analytic expressions of the torsional energy terms in terms of the folding angles are provided in
Sec.~\ref{sec:tors}. The fitting of the torsional terms of both 
Model 1 and 2 is described in Sec.~\ref{sec:fitting} completing the presentation of the derivation of the
new torsional force fields. Then, in Sec.~\ref{sec:zdis}, we present a test case, the deformation energy 
due to the vertical displacement of a carbon atom outside graphene's plane, as a
 first application of the proposed scheme and compare the prediction of Models 1 and 2 with DFT results. 
Finally, a summary and conclusions are given in Sect.~\ref{sec:conc}.

\section{Structures and DFT calculations\label{sec:struc}}

\subsection{Torsion angles in graphene}

\begin{figure}[!h]
    \centering
    \includegraphics[width=0.20\textwidth,clip]{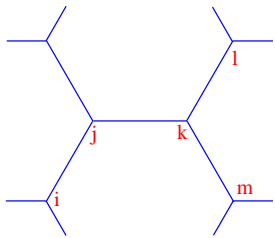}
    \caption{Graphene's honeycomb lattice with a few atomic positions labeled as i, j, k, l.}  
    \label{fig:dih}
\end{figure}

In Fig.~\ref{fig:dih}, we show a part of  the honeycomb 
structure of graphene and a few carbon-atom positions labeled as i, j, k, l, m.  The quadruple
(i, j, k, l), with 3 of these positions belonging to the same hexagonal ring and one to an 
adjacent is 
customary called ``trans" while the quadruple  (i, j, k, m), with all belonging to the same ring,
``cis". In the case that the structure is distorted and the atoms in the quadruple are no longer 
co-planar, we can define a torsion angle, which we label as (i-j-k-l),
as the dihedral angle of the planes i-j-k, j-k-l. The torsion angles can then be classified 
as ``trans" or ``cis" accordingly. 
The dihedral angle between two planes, e.g. i-j-k and j-k-l can be defined as the angle between
the vectors perpendicular to the planes. We assumed that the perpendicular vectors are pointing 
inwards for clockwise triples (like i-j-k) or outwards for anti-clockwise triples (like j-k-l). 
Under this assumption, torsional angles are in the range $[0,\pi]$ with ``cis" angles smaller
than $\pi/2$ and ``trans" angles larger than $\pi/2$.

\subsection{DFT results}

\begin{figure}[!tb]
\begin{tabular}{c}
     \includegraphics[width=0.4\textwidth,clip]{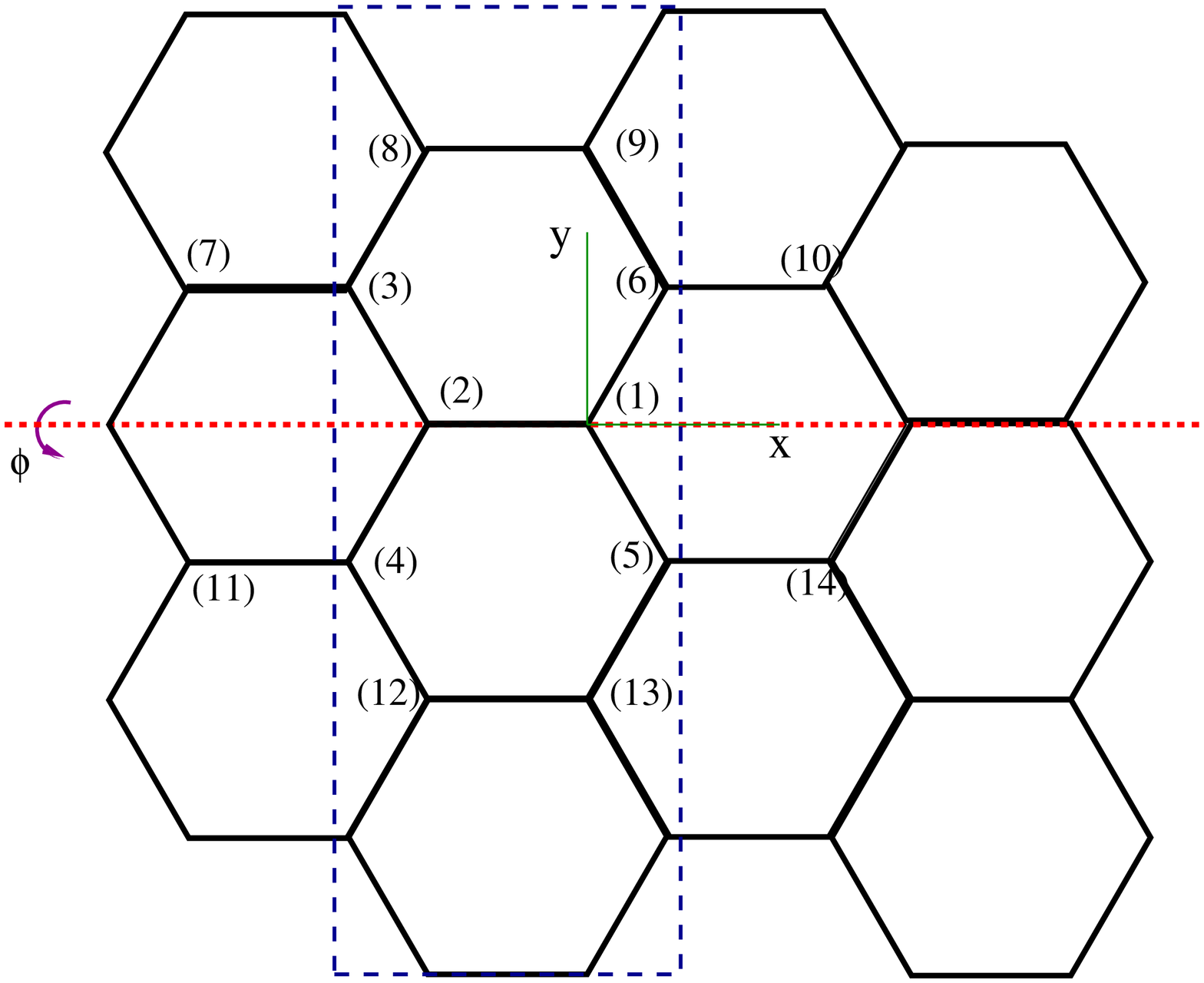}\\
     \includegraphics[width=0.4\textwidth,clip]{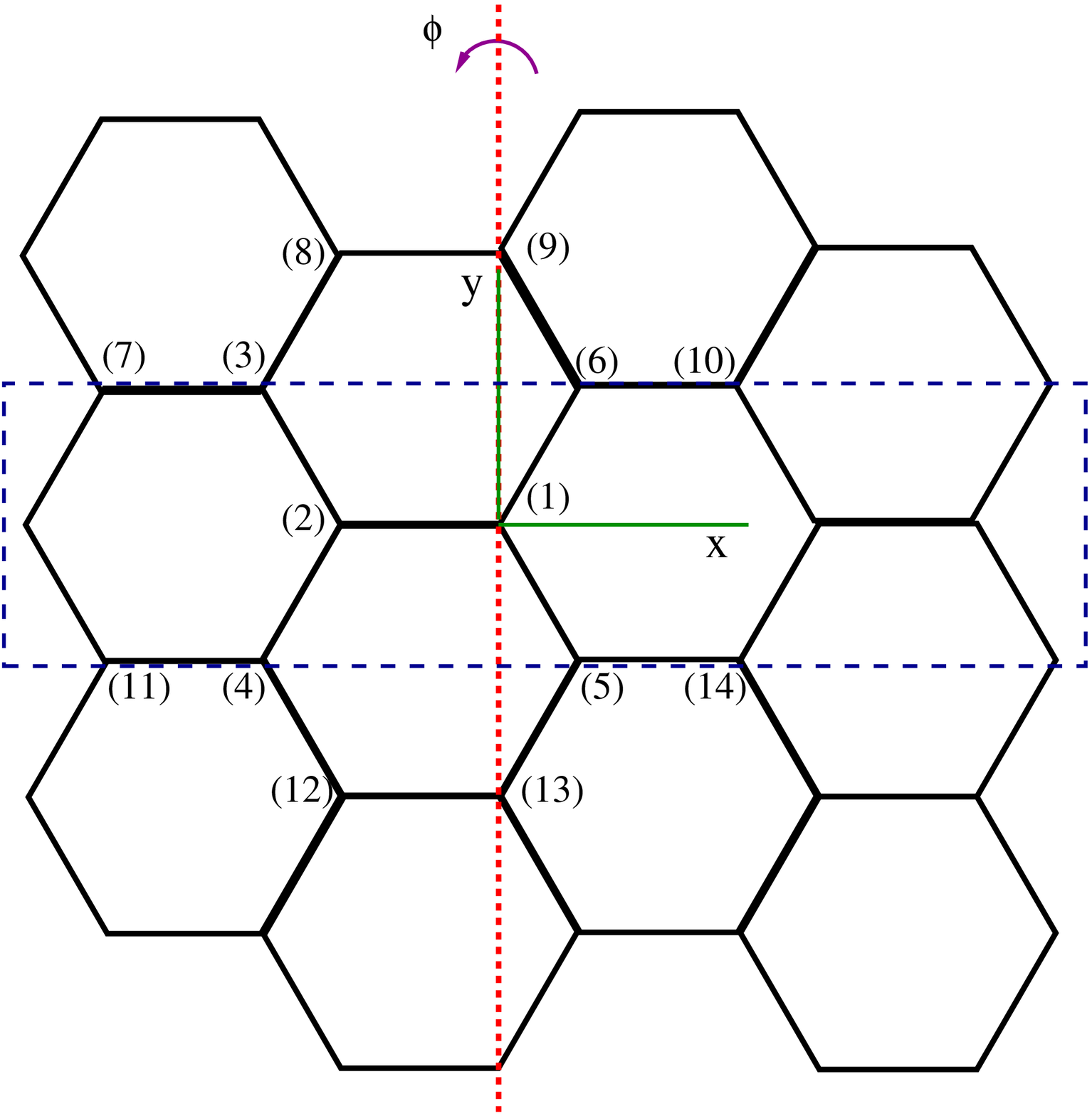}\\
\end{tabular}
\caption{ Part of graphene's structure with the armchair (top) and zig-zag (bottom) folding axes. We label several 
atoms that are mentioned in the text. The structure is periodic along the bending axis direction and 
part of the unit cell is shown in dashed blue line. The actual size of the unit cell along the direction vertical 
to the folding axis adopted in the DFT calculations is described in the text. 
}
\label{fig:struc}
\end{figure}

DFT calculations were performed for two distorted graphene structures, one that graphene is folded around an armchair
axis and one around a zig-zag. These structures are shown in Fig.~\ref{fig:struc} where
we label all atoms relevant to the present discussion. They are periodic along the folding axis while on
the vertical they are not.
The  folding angle around either the armchair or zig-zag axis is denoted by $\phi$. 
With  symbol $\theta$, we denote ``usual'' angles between carbon bonds (bending angles) and
with $\omega$ torsion angles as defined  in the previous subsection.

All calculations were performed with Quantum-Espresso periodic-DFT code \cite{QE-2009}, with the same
pseudopotential \cite{pseudo} as in Ref.~\onlinecite{kalosakas}. The wave-function and density plane-wave
cutoffs were chosen 40~Ry and 400~Ry respectively. The unit cell we chose is minimal in the periodic
direction (that of the folding axis) while, in the vertical, it is appropriately large to avoid edge-effects. 
Thus, the simulated structures are nanoribbons that are folded around their middle line direction. 
In the case of the armchair folding (Fig.~\ref{fig:struc}(top)), the vertical unit cell direction is such that 
neighbors up to the 5th in the vertical direction were included. In the case of the zig-zag folding that size is
long enough to include up to 8th neighbors. Thus, the unit cells contain 22 and 18 atoms for the armchair and
the zig-zag folding, respectively. In the reciprocal space, we used a mesh of
1$\times$24$\times$1, i.e. 24 points were assumed along the bending direction that the structure is periodic.

\begin{figure}
\includegraphics[width=0.48\textwidth,clip]{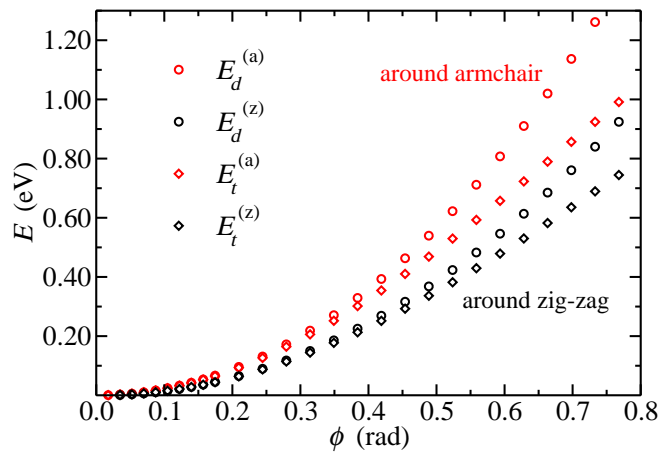} \\
\caption{ The total deformation energies, $E_d^{\rm (a)}$, $E_d^{\rm (z)}$, calculated with DFT
          and the total torsional energies, $E_t^{\rm (a)}$, $E_t^{\rm (z)}$, after removing angle bending terms, 
          as a function of the folding angle $\phi$. The indices (a), (z) correspond to the  armchair (red) and 
          zig-zag (black) folding cases, respectively.
        }
\label{fig:totalE}
\end{figure}

In Fig.~\ref{fig:totalE}, we show with filled circles the total deformation energy per unit cell along the folding direction,
as calculated by DFT, as a function of the folding angle $\phi$, for both armchair  and zig-zag folding actions,
$E_d^{\rm (a)}$ and $E_d^{\rm (z)}$, respectively. The total deformation energy per unit-cell is taken as the
energy difference  between the folded structure and the not folded one ($\phi=0$). 
Apparently, the two structure distortions due to the considered foldings are complex and consist of several individual
angle-bending and  torsional deformations. Note that bond lengths are not altered so there is no bond-stretching contribution 
in the total deformation energy. As we see in Fig.~\ref{fig:totalE}, the contribution from angle bending 
is significant for $\phi$ larger than 0.2~rad. In order to perform a fitting for the torsional terms alone, we first need 
to exclude angle-bending contributions from the total deformation energy. In order to do so, (i) we identify all 
angle-bending terms and express analytically their corresponding bending angles $\theta$ in terms of $\phi$ and then 
(ii) we remove the angle-bending terms using the analytic terms in the Ref.~\cite{kalosakas}. 
The residual, torsional energy per unit cell, when the contribution from angle-bending is subtracted, as a function of the out-of-plane
folding angle $\phi$ is shown  in Fig.~\ref{fig:totalE} (diamonds), for the two folding directions.
In order to fit an analytic expression to 
the torsional terms, we also have to (i) identify all the individual torsional terms that contribute 
for each of the zig-zag and armchair cases, and subsequently (ii) express the corresponding torsion
angles as functions of the folding angle, $\phi$. 
These steps are described below where we provide all necessary analytical expressions.

\section{Removing the angle-bending terms\label{sec:bend_rem}}

\subsection{For the folding around armchair axis}

The folding around the armchair direction (Fig.~\ref{fig:struc}, top)  alters 
two bond angles per unit cell, $(3\hat{2}4),(6\hat{1}5)$, which are equal. One can show that 
these angles, in terms of $\phi$, are given by 

\begin{equation}\theta^{\rm (a)}=2\arcsin\left({\sqrt{\frac{3}{8}}\sqrt{\cos\phi+1}}\right) \end{equation}

The angle-bending energy that one needs to remove from the total energy is  
\begin{equation}
U_{b}^{\rm (a)} = 2 V_b ( \theta^{\rm (a)}(\phi) )\,, 
\end{equation}
where $V_b(\theta)$ is the analytical expression for the angle bending given in Ref.~\cite{kalosakas}, i.e.
\begin{equation}  \label{abp}
V_b(\theta) = \frac{k}{2} \left( \theta-\frac{2\pi}{3} \right)^2 - \frac{k^\prime}{3} \left( \theta-\frac{2\pi}{3} \right)^3\,,
\end{equation}
with $k=7.0 \; {\rm eV}/{\rm rad}^2$ and $k^\prime=4 \; {\rm eV}/{\rm rad}^3$.

Removing these terms from the total deformation energies, $E_d^{\rm (a)}$ we find the 
total torsional energy
\begin{equation}
E_t^{\rm  (a)} = E_d^{\rm (a)} - U_{b}^{\rm (a)} \,,
\end{equation}

 shown in Fig.~\ref{fig:totalE}.

\subsection{For the folding around zig-zag axis}

Similarly, the folding around the zig-zag direction (Fig.~\ref{fig:struc}, (bottom)) 
affects two angles per unit cell, $(2\hat{1}6),(2\hat{1}5)$, that are also equal. 
In terms of $\phi$, these angles are given by 

\begin{equation}\theta^{\rm (z)}=2\arcsin\left(\frac{1}{2}\sqrt{2+\cos\phi}\right) \end{equation}

Again, the angle-bending energy that one needs to remove from the total energy is  
\begin{equation}
U_{b}^{\rm (z)} = 2 V_b ( \theta^{\rm (z)}(\phi) )\,, 
\end{equation}
we remove these terms from the total deformation energy and we find the total torsional energy,
\begin{equation}
E_t^{\rm  (z)} = E_d^{\rm (z)} - U_{b}^{\rm (z)} \,,
\end{equation}
shown in Fig.~\ref{fig:totalE}.

\section{Analytical expressions for the torsional terms\label{sec:tors}}

Here we provide analytical expressions, $U_{t}^{\rm (a)}(\phi)$, $U_{t}^{\rm (z)}(\phi)$, for the total torsional 
energies, as functions of $\phi$ that will contain parameters to be fitted so that these expressions 
reproduce as close as possible the $E_{t}^{\rm (a)}$, $E_{t}^{\rm (z)}$ points shown in Fig.~\ref{fig:totalE}.
To arrive to such analytical expressions we need first to identify all altered torsion angles (per unit cell) 
and express them in terms of the folding angle $\phi$. Then, $U_{t}^{\rm (a)}(\phi)$, $U_{t}^{\rm (z)}(\phi)$ will
be just the sum of all individual torsional terms that correspond to these altered torsional angles.

Regarding the individual torsional energy term, $V_t(\omega)$, two different functional forms would be considered.
The most frequently used formula, referred as Model 1 here, is 
\begin{eqnarray}
V_t(\omega) &=& \frac{1}{2} V_1 \left[ 1 + \cos  \omega  \right] \nonumber \\
 &+& \frac{1}{2} V_2 \left[ 1 - \cos (2 \omega ) \right] \label{eq:tors0} \,.
\end{eqnarray} 
An alternative model that we considered, which we call Model 2,
assumes a different fitting formula for cis or trans dihedral angles $\omega$
\begin{equation} 
\begin{array}{ll} 
V_t^{\rm (cis)}(\omega) = K_{\rm cis}   \left[ 1 - \cos (2 \omega ) \right]\,,  \\
V_t^{\rm (trans)}(\omega) = K_{\rm trans} \left[ 1 - \cos (2 \omega ) \right]\,,
\end{array}
\label{eq:tors2}
\end{equation} 
where either the first or the second expression is used for cis or trans torsion angles, respectively.
Below we use both Models 1 and 2 to fit their parameters to the obtained DFT results.

\subsection{For the folding around armchair axis}

Inspecting the Fig.~\ref{fig:struc} (top) we identify the following
torsion (dihedral) angles per unit cell that are altered by folding along the armchair axis
\begin{itemize}
  \item[$\bullet$]  2 trans dihedral angles, (5-1-2-3), (4-2-1-6), with
  \begin{equation} \omega_1^{\rm (a)}(\phi)=\arccos{\left(-\cos\phi\right)}\end{equation}
  \item[$\bullet$] 4 cis dihedral angles, (11-4-2-3), (14-5-1-6), (7-3-2-4), (10-6-1-5), with
  \begin{equation} \omega_2^{\rm (a)}(\phi)=\arccos{\left(\sqrt{3}\frac{1+\cos\phi}{\sqrt{9\sin^2\phi+6(1+\cos\phi)}}\right)}\end{equation}
  \item[$\bullet$] 4 trans dihedral angles, (12-4-2-3), (13-5-1-6), (8-3-2-4), (9-6-1-5), with
  \begin{equation} \omega_3^{(a)}(\phi)=\arccos{\left(-\sqrt{3}\frac{1+\cos\phi}{\sqrt{9\sin^2\phi+6(1+\cos\phi)}}\right)}\end{equation}
\end{itemize}

Through the angle expressions given above, the total torsional energy, $U_{t}^{\rm (a)}$ within the Model 1,
 becomes an analytic function of $\phi$:
\begin{eqnarray}
U_{t}^{\rm (a)}(\phi)  &= & 2 V_t (\omega_1^{\rm (a)}(\phi)) + 4 V_t ( \omega_2^{(a)}(\phi) ) \nonumber\\
 &+& 4 V_t ( \omega_3^{(a)}(\phi) ) 
\label{eq:Utors_arm}
\end{eqnarray}
where $V_t(\omega)$ is the individual torsional term Eq. (\ref{eq:tors0}).

For the Model 2, the corresponding expression of the total torsional energy $U_{t}^{\rm (a)}$ is
\begin{eqnarray}
U_{t}^{\rm (a)}(\phi) &=& 2 V_t^{\rm (trans)} (\omega_1^{(a)}(\phi)) + 4 V_t^{\rm (cis)} ( \omega_2^{(a)}(\phi) ) \nonumber \\
&+& 4 V_t^{\rm (trans)} ( \omega_3^{(a)}(\phi) ) 
\label{eq:Utors_arm2}
\end{eqnarray}
with $V_t^{\rm (cis)}$ and $V_t^{\rm (trans)}$ given by Eq. (\ref{eq:tors2}).

\subsection{For the folding around zig-zag axis}

Inspecting the Fig.~\ref{fig:struc} (bottom) we identify the following
dihedral angles per unit cell that are affected 

\begin{itemize}
  \item[$\bullet$]  2 cis dihedral angles, (3-2-1-6), (4-2-1-5), with
  \begin{equation} \omega_1^{\rm (z)}(\phi)=\arccos{\left(\sqrt{\frac{3}{\sin^2\phi+3}}\right)}\end{equation}\\
  \item[$\bullet$] 2 trans dihedral angles, (4-2-1-6), (3-2-1-5), with
  \begin{equation} \omega_2^{\rm (z)}(\phi)=\arccos{\left(-\sqrt{\frac{3}{\sin^2\phi+3}}\right)}\end{equation}\\
  \item[$\bullet$] 2 cis dihedral angles, (2-1-5-13), (2-1-6-9), with
  \begin{equation}\omega_3^{\rm (z)}(\phi)=\arccos{\left(\sqrt{\frac{3}{\sin^2\phi+3}}\cos\phi\right)}\end{equation}\\
  \item[$\bullet$] 2 trans dihedral angles, (2-1-5-14), (2-1-6-10), with
  \begin{equation} \omega_4^{\rm (z)}(\phi)=\arccos{\left(-\sqrt{\frac{3}{\sin^2\phi+3}}\cos\phi\right)}\end{equation}\\
\end{itemize}   

And the total torsional energy is given by
\begin{eqnarray}
U_{t}^{\rm (z)}(\phi) & = & 2 V_t (\omega_1^{\rm (z)}(\phi)) + 2 V_t ( \omega_2^{\rm (z)}(\phi) ) \nonumber \\
&+& 2 V_t ( \omega_3^{\rm (z)}(\phi)) + 2 V_t (\omega_4^{\rm (z)}(\phi))\,,
\label{eq:Utors_zig}
\end{eqnarray}
where $V_t$ is given by Eq.~(\ref{eq:tors0}) for Model 1.
In case of Model 2, the above formula becomes 
\begin{eqnarray}
\!\!U_t^{\rm (z)}(\phi) \!\!&=& \!\!2 V_t^{\rm (cis)} (\omega_1^{\rm (z)}(\phi)) + 2 V_t^{\rm (trans)} ( \omega_2^{\rm (z)}(\phi) ) \nonumber \\
\!\!\!\!&+&\!\!\! 2 V_t^{\rm (cis)} ( \omega_3^{\rm (z)}(\phi)) + 
2 V_t^{\rm (trans)} (\omega_4^{\rm (z)}(\phi)).  
\label{eq:Utors_zig2}
\end{eqnarray}

\section{Fitting procedure\label{sec:fitting}}

The total torsional energy data, $(\phi_i, E_{t,i}^{\rm (a)})$ and $(\phi_i, E_{t,i}^{\rm (z)})$,
shown in red and black diamonds in the Fig.~\ref{fig:totalE}, and the analytical (to be fitted) 
expressions $U_{\rm t}^{\rm (a)}$, $U_{\rm t}^{\rm (z)}$ given in the Eqs.~(\ref{eq:Utors_arm}) 
and (\ref{eq:Utors_zig}) for the Model 1 (or the Eqs.~(\ref{eq:Utors_arm2}) and (\ref{eq:Utors_zig2}) for the Model 2)
can be used to obtain the optimal parameters $V_1$ and $V_2$ (or $K_{\rm cis}$ and $K_{\rm trans}$) of the 
individual torsional terms so that $U_{\rm t}^{\rm (a)}$, $U_{\rm t}^{\rm (z)}$ reproduce the 
dependence of $E_{t}^{\rm (a)}(\phi)$ and $E_{t}^{\rm (z)}(\phi)$ as close as possible. 
For this purpose, adopting a standard procedure, we minimize an objective function 
$O(V_1,V_2)$ which is the equal-weighted sum of the square differences,
\begin{eqnarray}
O(V_1,V_2) &=& \sum_{i=1}^{\phi_i < \phi_{\rm max}} \left[ E_{t,i}^{\rm (a)} - U_{\rm t}^{\rm (a)}(\phi_i)\right]^2 \nonumber \\
&+& \sum_{i=1}^{\phi_i < \phi_{\rm max}} \left[E_{t,i}^{\rm (z)} - U_{\rm t}^{\rm (z)}(\phi_i)\right]^2 \,.
\end{eqnarray}
The sums in the above expression runs over all $i$ for which $\phi_i$ is smaller than an 
upper-limit angle $\phi_{\rm max}$.

For the Model 1, the fitted total torsional energies 
$U_{\rm t}^{\rm (a)}(\phi)$ and $U_{\rm t}^{\rm (z)}(\phi)$ given in the Eqs.~(\ref{eq:Utors_arm}) and (\ref{eq:Utors_zig})
depend on $V_1$ and $V_2$ through the dependence of the individual
terms $V_t$ of the Eq.~(\ref{eq:tors0}). 

In the case of Model 2, we are optimizing $K_{\rm cis}$ and $K_{\rm trans}$ parameters, and the expressions 
(\ref{eq:Utors_arm2}) and (\ref{eq:Utors_zig2}) are used instead and the individual terms 
$V_t$ are given by the Eq.~(\ref{eq:tors2}).

The choice of $\phi_{\rm max}$ is expected to affect the quality of fitting for small and large $\phi$.
We are interested in seeing whether the fitting parameters depend on $\phi_{\rm max}$ and, if so,
at what extend.

\subsection{Model 1: fitting Results for $V_1$, $V_2$}

\begin{figure*}[!t]
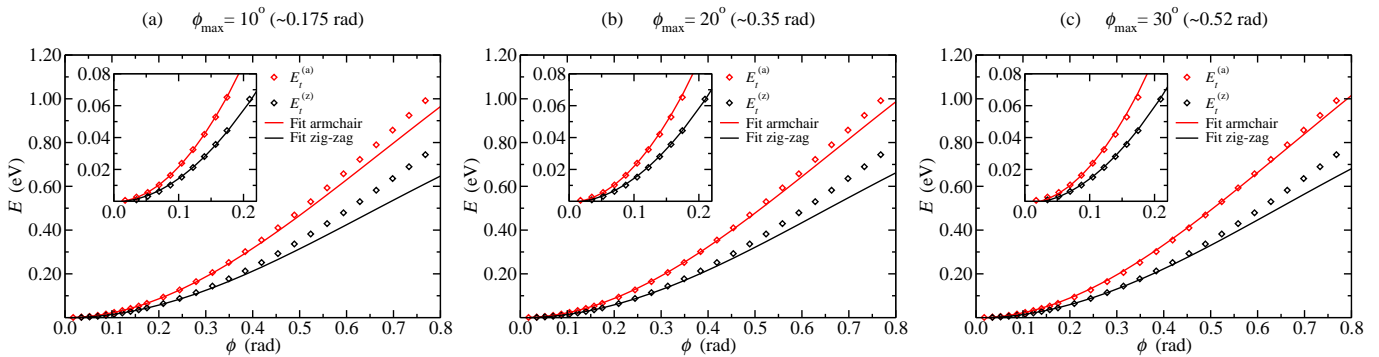

   \begin{tabular}{ccc}
   \includegraphics[width=0.33\textwidth,clip]{VFit10deg.eps} &
   \includegraphics[width=0.33\textwidth,clip]{VFit20deg.eps} &
   \includegraphics[width=0.33\textwidth,clip]{VFit30deg.eps} \\
   \end{tabular}
   \caption{{\bf Model 1:} Fit of the analytical expressions with the numerically derived data $E_t^{\rm (a)}$ and $E_t^{\rm (z)}$
   of the total torsional energy, for different choices of $\phi_{\rm max}$ equal to (a) $10^\circ$, (b) $20^\circ$, and (c) $30^\circ$.
   In the inset we zoom in the region of small angles $\phi$.} 
   \label{fig:Vfit}
\end{figure*}

We performed fitting of $V_1$, $V_2$ of Eq.~(\ref{eq:tors0}), for three different 
$\phi_{\rm max}$ values: 10$^\circ$, 20$^\circ$ and 30$^\circ$.
The optimal parameters $V_1$ and $V_2$ given in table~\ref{tab:optV}.

\begin{table}[!h]
\begin{tabular}{|l|c|c|}
\hline
$\phi_{\rm max}$   &    $V_1$ (eV) & $V_2$ (eV) \\
\hline\hline
   10              & -0.00013 & 0.221 \\
   20              & -0.00017 & 0.226 \\
   30              & -0.00035 & 0.233 \\
\hline
\end{tabular}
\caption{{\bf Model 1:} Optimal fitting parameters $V_1$ and $V_2$ for Model 1, for the three different values of 
$\phi_{\rm max}$ examined.} \label{tab:optV}
\end{table}

In Fig. \ref{fig:Vfit}, we show, the total torsional energies, $E_t^{\rm (a)}$ and 
$E_t^{\rm (z)}$ and the fitted lines for these three values of $\phi_{\rm max}$. 
As we see, the fitting is better for small values of $\phi$ and deteriorates as $\phi$ increases.
In all cases, it reproduces the armchair data in closer agreement than
the zig-zag, i.e. for a larger range of $\phi$.
For $\phi_{\rm max}=10^\circ$, there is a satisfactory agreement for values of 
$\phi$ up to 0.55~rad for the armchair case and 0.4~rad for the zig-zag case. For  
$\phi_{\rm max}=20^\circ$, the range of satisfactory agreement increases roughly up to 0.65~rad 
and 0.45~rad for the armchair and zig-zag cases respectively. Finally for 
$\phi_{\rm max}=30^\circ$, 
the agreement range increases further up to 0.7~rad and 0.5~rad, respectively. Although by 
increasing $\phi_{\rm max}$, the range of satisfactory agreement also increases, this is at the
cost of the agreement for smaller angles. As the code is trying to fit better at larger values
of $\phi$ the quality for smaller angles deteriorates. This deterioration, however, is rather 
small as we observe in the insets of the Figs.~\ref{fig:Vfit}(a), (b), (c) where we zoom in that 
region. 

\begin{figure}[!h]
 \includegraphics[width=0.48\textwidth,clip]{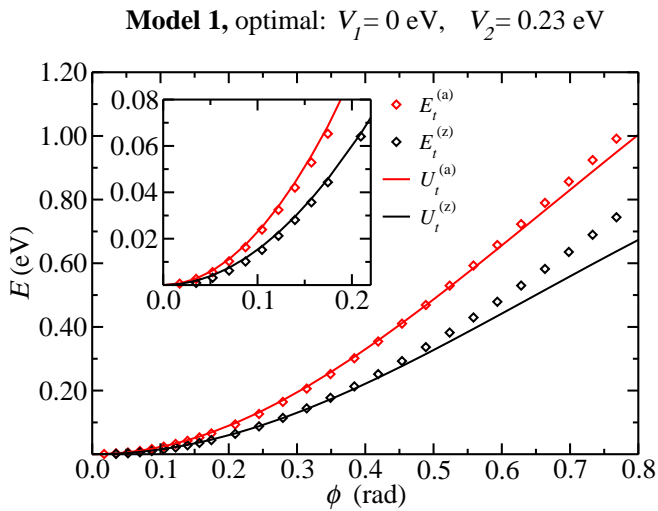}
 \caption{ The analytical torsional energies considering Model 1, for the folding around the armchair (a) 
     and zig-zag (z) axes, $U_t^{\rm (a)}$ and $U_t^{\rm (z)}$, Eqs.~(\ref{eq:Utors_arm}) and (\ref{eq:Utors_zig})
     respectively, using the proposed parameters $V_1=0$ eV and $V_2$=0.23 eV (lines), compared to the numerically derived data
     $E_t^{\rm (a)}$ and $E_t^{\rm (z)}$ (points). The inset zooms in small values of $\phi$ to illustrate the quality of the fit in that region.  }
 \label{Vopt}
\end{figure}

However, in general, for angles $\phi$ up to 0.4~rad (22$^\circ$) corresponding to energies ~0.2 to ~0.3 eV,
all fittings are satisfactory. On the other hand, as we see in Table~\ref{tab:optV},
the fitted values of $V_1$ and $V_2$ are not so sensitive to the
value of $\phi_{\rm max}$: $V_1$ remains close to 0 while $V_2$ is in the range 0.22-0.23 eV. 
In addition, the large value of $V_2$, i.e. 0.23~eV, obtained for $\phi_{\rm max}=20^\circ,30^\circ$ 
performs better for larger angles, up to 0.5 rad ($\sim$30$^\circ$), corresponding to energies of 
0.4-0.5~eV, while on the other hand the fitted results in the region of small $\phi$ remain 
satisfactory. These considerations suggest that it is quite reasonable to adopt as optimal $V_1=0$ and $V_2=0.23$~eV
and our proposed torsional potential has the simple form
\begin{equation} 
    V_t(\omega)=\frac{1}{2}V_2\left(1-\cos(2\omega)\right),\quad V_2=0.23\; {\rm eV}\,. \label{eq:opt}
\end{equation}
This potential is shown in Fig.~\ref{Vopt} together with $E_t^{\rm (a)}$ and $E_t^{\rm (z)}$.
%However, one has to have in mind that for simulations of situations with small out of plane deformations
%a slightly smaller value of $V_2$ might be slightly better.

\subsection{Model 2: fitting results for $K_{\rm cis}$, $K_{\rm trans}$}

\begin{figure*}[t]
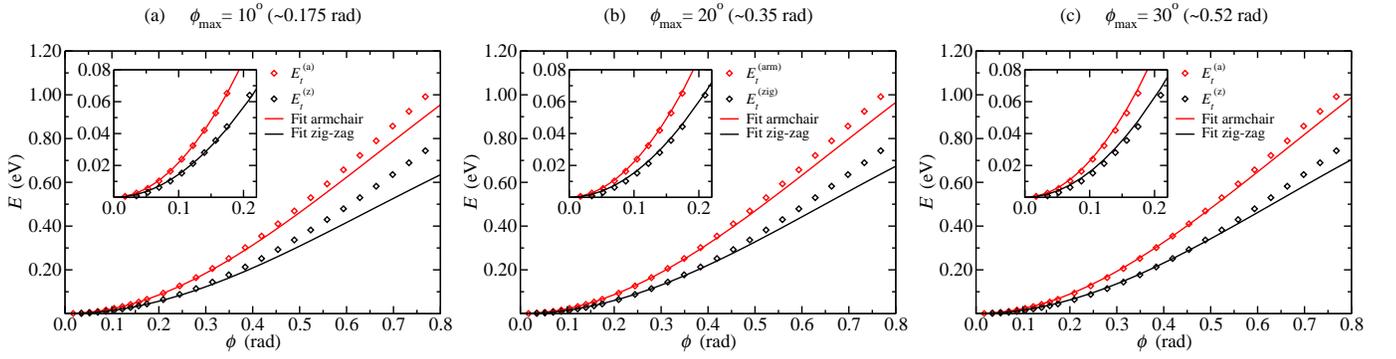

\centering
\begin{tabular}{ccc}
\includegraphics[width=0.33\textwidth,clip]{KFit10deg.eps} &
\includegraphics[width=0.33\textwidth,clip]{KFit20deg.eps} &
\includegraphics[width=0.33\textwidth,clip]{KFit30deg.eps} \\
\end{tabular}
\caption{ {\bf Model 2:} Fit of the analytical expressions with the numerically derived data $E_t^{\rm (a)}$ and $E_t^{\rm (z)}$
   of the total torsional energy, for different choices of $\phi_{\rm max}$ equal to (a) $10^\circ$, (b) $20^\circ$, and (c) $30^\circ$.
   In the inset we zoom in the region of small angles $\phi$.  } 
   \label{fig:Kfit}
\end{figure*}

\begin{figure}[!h]
  \centering
  \includegraphics[width=0.48\textwidth,clip]{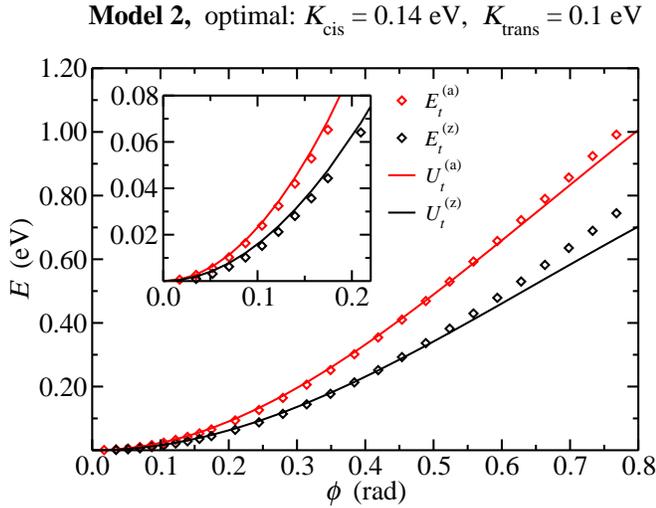}
  \caption{ The analytical torsional energies considering Model 2, for the folding around the armchair (a) 
        and zig-zag (z) axes, $U_t^{\rm (a)}$ and $U_t^{\rm (z)}$, Eqs.~(\ref{eq:Utors_arm2}) and (\ref{eq:Utors_zig2}) respectively,
        using the proposed parameters $K_{\rm cis}$=0.14 eV, $K_{\rm trans}$=0.10 eV, compared to the data $E_t^{\rm (a)}$ and
        $E_t^{\rm (z)}$. The inset zooms in small values of $\phi$ to illustrate the quality of the fit in that region.   }
  \label{Kopt}
\end{figure}

As in the previous section, we performed the fitting of $K_{\rm cis}$ and $K_{\rm trans}$
of Eq.~(\ref{eq:tors2}) for the same values of $\phi_{\rm max}$, i.e. 10$^\circ$, 20$^\circ$ and 30$^\circ$.  
The optimal parameters $K_{\rm cis}$ and $K_{\rm trans}$ for each of these cases are given in 
Table~\ref{tab:optK}.
In Fig.~\ref{fig:Kfit}, we show the fitting lines for the armchair and zig-zag folding cases 
for all three values of $\phi_{\rm max}$
compared to the data points for the total torsional energy per unit cell, $E_t^{\rm (a)}$ and 
$E_t^{\rm (z)}$. The fitting quality is quite similar to that of the previous section. Again,
 for $\phi_{\rm max}=20^\circ$ and $30^\circ$, the quality improves for larger 
values of $\phi$ and at the same time the fit for smaller $\phi$ does not deteriorate substantially.
Thus, we propose a rounded optimal set 
$K_{\rm cis}$=0.14 eV and $K_{\rm trans}$=0.10 eV which is close to the values obtained for 
$\phi_{\rm max}=20^\circ$ and $30^\circ$. In Fig.~\ref{Kopt}, we show the torsional energy 
obtained with Model 2 and these values for $K_{\rm cis}$ and $K_{\rm trans}$ compared with the
data $E_t^{\rm (a)}$ and $E_t^{\rm (z)}$.

\begin{table}[!h]
\begin{tabular}{|l|c|c|}
\hline
$\phi_{\rm max}$   &    $K_{\rm cis}$ (eV) & $K_{\rm trans}$ (eV) \\
\hline\hline
   10              & 0.104 & 0.112 \\
   20              & 0.134 & 0.096 \\
   30              & 0.150 & 0.090 \\
\hline
\end{tabular}
\caption{{\bf Model 2:} Optimal fitting parameters $K_{\rm cis}$ and $K_{\rm trans}$, 
for the three different values of $\phi_{\rm max}$ examined.}  \label{tab:optK}
\end{table}

To summarize, for the Model 2, we propose
{\small\begin{equation} 
\begin{array}{ll} 
V_t^{\rm (cis)}(\omega) = K_{\rm cis}   \left[ 1 - \cos (2 \omega ) \right], \quad K_{\rm cis}=0.14 \mbox{\ eV} \,,  \\
V_t^{\rm (trans)}(\omega) = K_{\rm trans} \left[ 1 - \cos (2 \omega ) \right], \mbox{} K_{\rm trans}=0.1 \mbox{\ eV}\,,
\end{array}
\label{eq:tors3}
\end{equation} }
where either the first or the second expression is used depending on whether the
torsional angle $\omega$ is cis or trans.

\subsection{Comparison of the two models}

\begin{figure}[!h]
\centering
\includegraphics[width=0.48\textwidth,clip]{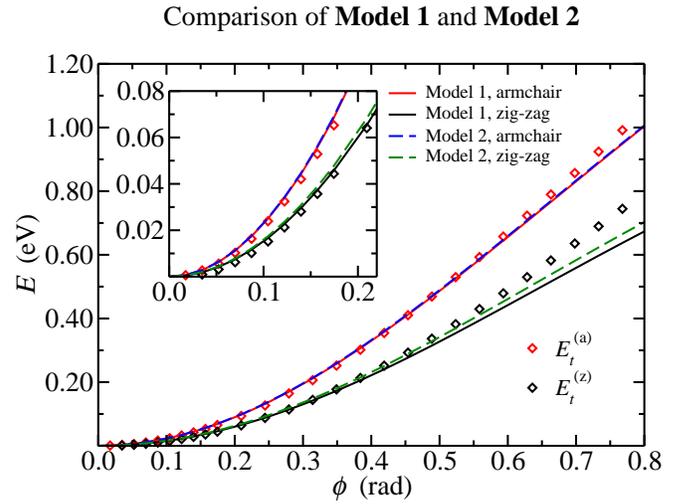}
\caption{Comparison between the fitting of the two models: Model 1 with values $V_1$=0 eV, $V_2$=0.23 eV and Model 2
with values $K_{\rm cis}$=0.14 eV, $K_{\rm trans}$=0.10 eV.}
\label{All}
\end{figure}

In Fig.~\ref{All} we show the torsional energies per unit cell, for both fitting forms of Models 1 and 2,
for the  case of the optimal parameters we arrived at.  We notice that the two models
are of the same quality. They almost coincide for the armchair case, while for the
zig-zag, the Model 2 is slightly better for larger $\phi$'s and the Model 1 marginally better for smaller.
The differences however are not significant for $\phi$'s up to 0.5 rad.

Note that the obtained optimal parameters of Model 2, $K_{\rm cis}$=0.14~eV and $K_{\rm trans}$=0.10~eV,
are close to each other, indicating that a single parameter with value the average of them 
would offer a reasonable description. Moreover, this average value is almost equal to $ V_2 /2$.
Thus, it is rather unnecessary to assume different parameters for {\it cis} and {\it trans} dihedral angles 
and, to keep things as simple as possible, the simple form of the Eq.~(\ref{eq:opt}), is quite sufficient to describe
all torsional distortions. Therefore, the Model 1 of Eq.~(\ref{eq:opt}) is our proposed one, containing a
single parameter $V_2$. We should mention that our modeling describes accurately the energy of torsional
angles $\phi$ up to 0.5 rad which is already a sufficiently large value, corresponding to rather unphysical structural 
deformations.

\section{Application to the vertical displacement of a carbon atom in graphene\label{sec:zdis}}

In order to test the accuracy of the proposed parameters for the torsional terms, 
we consider the deformation energy of graphene due to a vertical, out-of-plane displacement 
of a single carbon atom. We consider that apart from the vertically displaced atom, all other atoms remain
fixed at their equilibrium positions within graphene's plane. The task is to compare 
the deformation energy obtained by the present potential, along with the in-plane force field of Ref.~\cite{kalosakas},
with that obtained by DFT calculations (using the same method that was used to produce the data discussed above). 

The process of moving a carbon atom vertically outside graphene's plane is described by a deformation 
energy consisting of all kinds of individual terms, i.e. bond-stretching, 
angle-bending and of course torsional terms. 

\begin{figure}[!h]
    \centering
    \includegraphics[width=0.3\textwidth,clip]{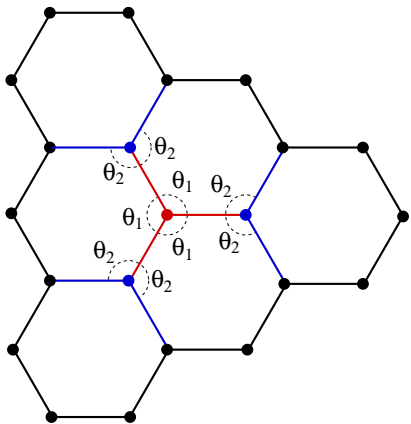}
    \caption{A vertical displacement (normal to the page) of a C atom (marked red)
    results in contributions to the total deformation 
    energy from bond-stretching (elongation of red bonds), angle-bending (altering $\theta_1$ 
    and $\theta_2$ angles), and torsional terms (twisting around red and blue bonds).  
    }\label{bonds}
\end{figure}

Concerning the bond-stretching terms, the vertical movement of a carbon atom at a displacement $z$ 
over the plane, alters only the three bonds of that atom (see Fig.~\ref{bonds}). If the length 
of these bonds at $z$=0 is $d$ their altered length $d'$ becomes 
\begin{equation}
  d'(z)=\sqrt{d^2+z^2}\label{newbond}
\end{equation}

At the same time, two different kinds of angle-bending terms appear corresponding to: (i) the three angles, $\theta_1$,
between the atom's bonds (marked in red in Fig.~\ref{bonds}) and (ii) the six angles, $\theta_2$ between these bonds
and the bonds marked in blue in Fig.~\ref{bonds}. These angles can be expressed in terms of the displacement $z$ as
\begin{equation}
\theta_1(z)=\arccos\left(\frac{2z^2-d^2}{2(d^2+z^2)}\right)\label{newtheta}
\end{equation}
and 
\begin{equation}
\theta_2(z)=\arccos\left(-\frac{d}{2\sqrt{d^2+z^2}}\right)\label{newphi}
\end{equation}

Finally, several torsional terms also contribute. There are rotations around the 3 bonds 
of the displaced atom and its first neighbors (marked in red in Fig.~\ref{bonds}) as well as 
rotations around the 6 bonds of the first neighbors and the second neighbors. These
rotations correspond to the following torsional angles
\begin{itemize}
\item 6 cis dihedral angles around the bonds marked in red in Fig.~\ref{bonds} given by
      \begin{equation}
          \omega_{\rm cis}^{(1)}(z)=\arccos\left(\frac{\frac{3}{4}}{\sqrt{\left(\frac{z}{d}\right)^2+
          \frac{3}{4}}\sqrt{3\left(\frac{z}{d}\right)^2+\frac{3}{4}}}\right)\label{last1}
      \end{equation} 
\item 6 trans dihedral angles around the bonds marked in red in Fig.~\ref{bonds} given by
      \begin{equation}
          \omega_{\rm trans}^{(1)}(z)=\arccos\left(\frac{-\frac{3}{2}\left(\frac{z}{d}\right)^2-
          \frac{3}{4}}{\sqrt{\left(\frac{z}{d}\right)^2+\frac{3}{4}}\sqrt{3\left(\frac{z}{d}
          \right)^2+\frac{3}{4}}}\right)
      \end{equation}
\item  6 cis dihedral angles around the bonds marked in blue in Fig.~\ref{bonds} given by
      \begin{equation}
          \omega^{(2)}_{\rm cis}(z)=\arccos\left(\frac{1}{\sqrt{\frac{4}{3}\left(\frac{z}{d}
          \right)^2+1}}\right)
       \end{equation}
\item 6 trans dihedral angles around the bonds marked in blue in Fig.~\ref{bonds} given by
       \begin{equation}
           \omega^{(2)}_{\rm trans}(z)=\arccos\left(\frac{-1}{\sqrt{\frac{4}{3}\left(\frac{z}{d}
           \right)^2+1}}\right)\label{last}
       \end{equation}
\end{itemize}

\begin{figure}
\includegraphics[width=0.45\textwidth,clip]{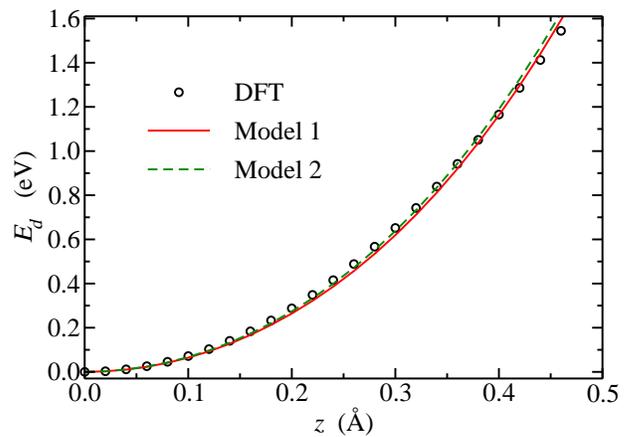} \\
\caption{The deformation energy, $E_d$, written in analytical form in Eq.~(\ref{eq:E_z}), due to the vertical out-of-plane
displacement of a single carbon atom in graphene as a function of the vertical displacement $z$,
calculated with the present potential and compared to DFT results.}
\label{fig:vertical}
\end{figure}

Then, the deformation energy is given by
\begin{eqnarray}
  E_d(z) &=& 3 V_s (d'(z)) + 3 V_b (\theta_1(z)) + 6 V_b (\theta_2(z))\nonumber \\ 
         &+& 6 V_t(\omega_{\rm cis}^{(1)}(z)) + 6 V_t(\omega_{\rm trans}^{(1)}(z)) \nonumber\\
         &+& 6 V_t(\omega_{\rm cis}^{(2)}(z)) + 6 V_t(\omega_{\rm trans}^{(2)}(z))\label{eq:E_z} \,,
\end{eqnarray}
where $V_s (d)$, $V_b (\theta)$ are respectively the bond stretching and angle bending terms 
given in Ref.~\cite{kalosakas}. $V_t$ is the individual torsional term of Eq.~(\ref{eq:opt})
for Model 1, while for Model 2 it should be replaced by $V_t^{\rm (cis)}$ or $V_t^{\rm (trans)}$
of Eq.~(\ref{eq:tors3}) for the two $\omega_{\rm cis}$ and the two $\omega_{\rm trans}$
respectively.
In Eq.~(\ref{eq:E_z}), the deformation energy, $E_d$, becomes an analytic function of $z$ through the
explicit dependence on $z$ of the bonds $d'$, the angles $\theta_1$, $\theta_2$ and the dihedral angles
$\omega_{\rm cis}^{(1)}$, $\omega_{\rm trans}^{(1)}$, $\omega_{\rm cis}^{(2)}$, $\omega_{\rm trans}^{(2)}$.
 
In Fig.~\ref{fig:vertical},  we show $E_d$ in comparison with the corresponding DFT results. 
As we see, both models perform equally well and the error does not exceed 0.05~eV for the 
deformation range shown.  There are small differences in their 
agreement with DFT, for example Model 2 seems slightly better for indermediate-size 
displacements while Model 1 for larger ones, however, these small differences are 
insignificant validating our preference for Model 1 on the basis of its simplicity. 

\section{Conclusion\label{sec:conc}}

In summary, we present a simple torsional force field for graphene and other sp$^2$ carbon nanostructures.
To obtain this potential we performed two sets of DFT calculations by folding two different graphene nanoribbon
 structures around their middle line. The first set of calculations concern the folding of an armchair nanoribbon 
around its middle line, an armchair direction. The second concerns the folding of a zig-zag nanoribbon around its 
middle line which is a zig-zag direction. From the deformation energies we isolated the ``pure'' torsional
contribution by removing angle bending terms with the use of our previously proposed angle bending terms\cite{kalosakas}.
The purified torsional deformation energy was then fit two different analytic forms with two parameters each that we
call Models 1 and 2. The first (Model 1) is that of Eq.~(\ref{eq:tors0}) and does not distinguish torsional angles, while
the second (Model 2) of Eq.~(\ref{eq:tors2}) treats differently ``cis'' and ``trans'' torsional angles. We found that 
the form of Model 1 reduces to one parameter form, see Eq.~(\ref{eq:opt}), which was found to be an average of the ``cis'' 
and ``trans'' terms of Model 2 which differ very little from each other, see Eq.~(\ref{eq:tors3}). That suggests that the 
use of two different terms is redundant and the single term of Model 1 suffices at a reasonable level or accuracy. We found 
that both models reproduce accurately the torsional deformation energy of graphene nanoribbons due to the folding we 
considered up to $\theta_{\rm max}$ of the order of $30^o$~($\approx 0.5$~rad).

As an additional validation test we considered the case of the deformation energy due to the vertical displacement of a 
single C atom outside of graphene's plane. For this task we used the torsional terms of either Model 1 or 2 combined with 
the bond stretching and angle bending terms of Ref.~\onlinecite{kalosakas}. For all terms in this formula we provide 
analytic expressions in terms of the displacement $z$. We found that both models perform equally well in this case with errors
not exceeding 0.05~eV for a relatively large range of $z$, up to 0.4-0.5~\AA. The good performance of both Models in this case
validates our choice for the simpler Model 1. 

The torsional force field presented here, in combination with the bond stretching and angle bending terms of Ref.~\onlinecite{kalosakas},
which were also fitted to DFT results using the same density functional approximation provide a complete, accurate, but simple in form
atomistic potential, which is computationally efficient due to its simplicity. Thus, we expect that 
it will be proven a very useful tool for large scale atomistic simulation of graphene and other sp$^2$ nanostructures.   

{\it Acknowledgements:}
We acknowledge helpful discussions with E. N. Koukaras.
The research leading to the present results has received funding from Thales project
``GRAPHENECOMP'', co-financed by the European Union (ESF) and the Greek Ministry of Education 
(through ΕΣΠΑ program). NNL  acknowledges support from the Hellenic Ministry of Education/GSRT (ESPA), through 
"Advanced Materials and Devices" program (MIS:5002409) and EU H2020 ETN project ‘Enabling Excellence’ Grant 
Agreement 642742.

  \end{document}